\documentclass[%
 aip,
 amsmath,amssymb,
reprint,%
]{revtex4-1}

\usepackage{graphicx}
\usepackage{dcolumn}
\usepackage{subcaption}
\usepackage{comment}
\usepackage{caption}
\usepackage{hyperref}
\hypersetup{
    colorlinks=true,
    linkcolor=blue,
    citecolor =blue,
    urlcolor=blue,
    }
\usepackage[mathlines]{lineno}

\usepackage[utf8]{inputenc}
\usepackage[T1]{fontenc}
\usepackage{txfonts}
\usepackage{etoolbox}

\newcommand{\eref}[1]{Eq.~(\ref{#1})}

\newcommand{\figref}[1]{Fig.~\ref{#1}}

\usepackage{verbatim}

\makeatletter
\def\@email#1#2{%
 \endgroup
 \patchcmd{\titleblock@produce}
  {\frontmatter@RRAPformat}
  {\frontmatter@RRAPformat{\produce@RRAP{*#1\href{mailto:#2}{#2}}}\frontmatter@RRAPformat}
  {}{}
}%
\makeatother

\begin{document}

\preprint{AIP/123-QED}

\title{Strong actuation and optomechanical application of mass loaded membranes}

\author{Joe Depellette}
\author{Ewa Rej}
\author{Richa Cutting}
\author{Mika A. Sillanpää}
\email[Electronic mail: ]{joe.depellette@aalto.fi; mika.sillanpaa@aalto.fi}

\affiliation{$^1$QTF Centre of Excellence, Department of Applied Physics, Aalto University, FI-00076 Aalto, Finland}

\date{\today}

\date{\today}

\begin{abstract}

An increasing number of studies are moving towards the combination of quantum mechanics and gravity, where studying gravity from a very small source mass is a viable starting point. Preparing for such experiments, investigations of weak gravitational forces have employed mechanical resonators to detect time-dependent gravitational forces from actuated source masses. Here, we demonstrate a source mass approach which utilizes capacitive actuation of a 1\,mg gold sphere embedded on a silicon nitride membrane, rather than piezoelectric or motorized actuation. The design simultaneously provides a method for microwave optomechanical implementation by coupling the membrane position to the electromagnetic mode of a 3D cavity. The cavity quality factor is not significantly compromised by electromagnetic leakage to the actuation electrode, allowing DC and kilohertz AC voltages to be introduced in the region where electric fields are strongly concentrated. We measure over 700\,nm of driven oscillation amplitude and more than ten percent tunability in the mechanical resonance frequency of the loaded membrane, giving the potential to match the oscillations to the frequency range of a detector in future experiments. An optomechanical readout is demonstrated by measuring the cavity resonance at cryogenic temperatures, while room temperature measurements provide complimentary understanding of the mechanisms which influence the mechanical response, including repulsive contact due to collisions within the device.

\end{abstract}

\maketitle

\section{Introduction}


The utilization of mechanical oscillators in force or acceleration sensing is not a new phenomenon, with pendulums demonstrating small variations in the Earth’s gravitational field even before the publication of Newton’s law of universal gravitation\cite{lenzen1965development}. Over the centuries that followed, systems involved in force detection have become increasingly sensitive, and quite often, significantly smaller. In the modern day, mechanical resonators of the micro and nano scale are combined with electrical interfaces to create micro/nanoelectromechanical systems (M/NEMS). These provide vital accelerometers for much of the electronics used in society. M/NEMS devices can also be used to measure electric\cite{horenstein2001micro,peng2006design,bahreyni2008analysis,kobayashi2008microelectromechanical,yang2013design,kainz2018distortion,kareekunnan2021revisiting,zhu2025non}, magnetic\cite{nan2013self,li2017ultra,xu2024subpicotesla}, and gravitational\cite{middlemiss2016measurement,middlemiss2018microelectromechanical,westphal2021measurement,fuchs2024measuring} fields with high sensitivity. 

Unlike the pendulum measurements of the 17th century, there is now a noteworthy demand for the gravitational fields in basic studies to be of a manufactured origin. In particular, studying gravity produced by a very small source mass in the milligram range or below offers a viable route towards charting the transition between quantum mechanics and gravity. With technology progressing, these studies are likely to yield significant results in the not so distant future. The present study introduces a device capable of producing a controllable gravitational field from a 1\,mg source mass. The particular applications link to a proposed experiment to measure gravitational forces in a microwave optomechanical setup\cite{liu2021gravitational}, with a route towards preparing self-gravitating quantum-mechanical systems.

\section{Device}



The compromise between the mass of a gravitational source and its various mechanical properties is an important consideration when designing a device. Following the recent trends of small gravitational force measurements, highly sensitive detectors measure the response due to a time-dependent gravitational field. The device carrying the source mass must therefore be able to produce a modulated field by coherently driving oscillations of the source mass position, while having a frequency tuning capability in order to match the resonance of a detector. For this, we use a high-stress silicon nitride (SiN) membrane with a capacitive coupling to a nearby gate electrode.


Much like detectors making use of mechanical resonance, the source mass also benefits from being a resonator by having an enhanced amplitude when driven at the correct frequency. Tuning the resonance with a DC bias is limited by the pull-in voltage, and so the unbiased resonances of a source and detector must be within a certain frequency range to be compatible.


A source must be massive enough to produce a detectable gravitational force in the first place. A MEMS device such as a bare SiN membrane requires mass loading in order to fulfill this criteria. For force or acceleration detectors, higher frequencies provide benefits in mitigating $1/f$ noise of both acoustic and electrical origin. Higher frequencies are also beneficial for reaching lower thermal noise when approaching the quantum regime of vibrations. On the contrasting side, sensing of gravity equates to acceleration sensing, which suffers at higher frequencies. We consider the kilohertz range studied here, with milligram masses involved, to be a sweet spot in meeting conflicting requirements for both source and test masses. With the future aim of preparing a source mass in a state which exhibits macroscopic quantum phenomena, such as ground-state cooling or dissipative quantum squeezing of the oscillator, our mass-loaded membranes present an opportunity to produce a significantly large gravitational field originating from a nonclassical source. The combination of gravitational and quantum effects in an experimental context is a necessity for the future of fundamental studies of quantum gravity.

\begin{figure*}
    \centering
    \includegraphics[width=0.9\linewidth]{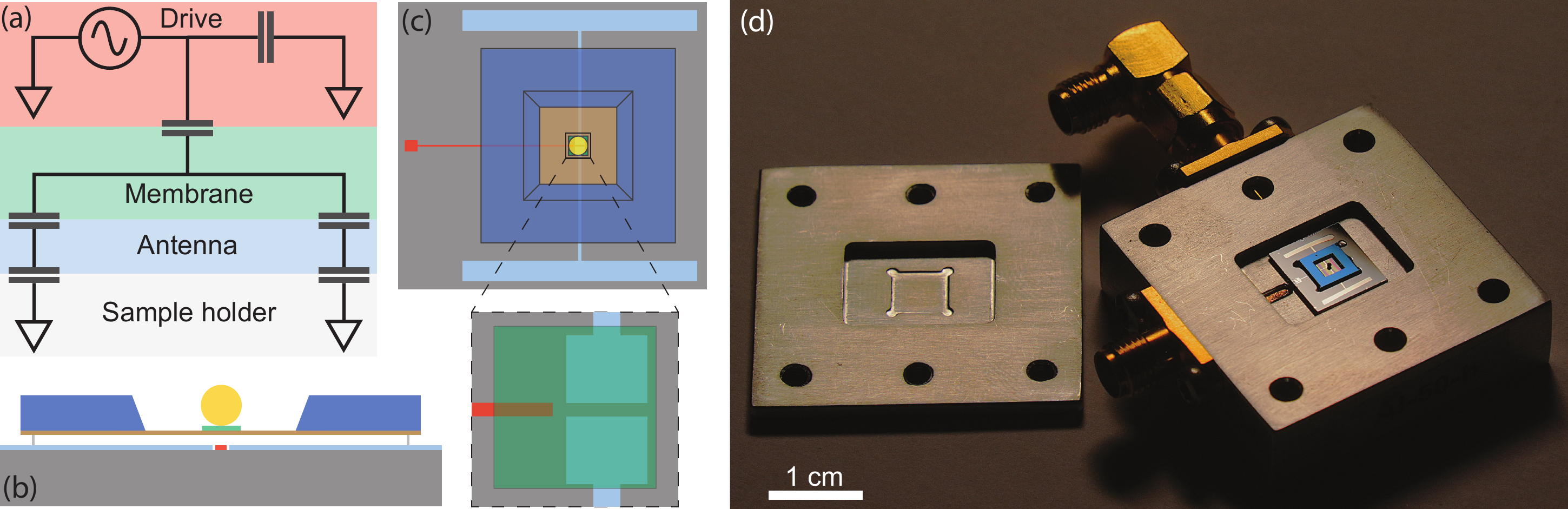}
    \caption{Flipchip device schematic. (a)~Equivalent circuit diagram used to simulate the voltage division within the system. (b)~Side view and (c)~top-down view of the device design. (d)~Photograph of the device mounted in the 3D cavity sample holder.}
    \label{fig: device1}
\end{figure*}

\subsection{Mechanical properties}

Detailed in \figref{fig: device1}, the device consists of a 2\,mm square, 100\,nm thick, metallized SiN membrane suspended from a silicon frame, with a gold sphere of 500\,$\mu$m diameter and mass $M = 1.3$\,mg glued to the center of the membrane using epoxy. A typical unloaded membrane of this type has a mechanical resonance frequency of around 200\,kHz, determined by the geometry and stress. The mass loading significantly decreases the resonance frequency. The loaded frequency, denoted by $\omega_\mathrm{m}$, arrives in the range $\omega_\mathrm{m}/2\pi \approx 2$\,kHz, additionally being affected by the size of the glue drop. A flipchip is formed by gluing the membrane frame to a separate antenna chip, where the gap size denoted below as $d$, typically of the order 1\,$\mu$m, is determined by a combination of dust and distortions in the shape of the frame. The addition of the gold sphere does not significantly deflect the membrane equilibrium due to its weight.

The silicon antenna chip contains lithographically patterned aluminum structures, one of which is a gate electrode used for electrostatic tuning and actuation. Simply driving a source mass at kilohertz frequencies does not require this somewhat complicated scheme; one could bypass the need for a membrane altogether and directly attach the mass to a vibrating piezoelectric material. The true benefit of the presented flipchip device comes from its ability to form part of an optomechanical system, with two patterned antennas on the silicon chip allowing the membrane to couple to the electromagnetic mode of a 3D cavity. The coupling occurs via the displacement-dependence of the capacitor formed by the vacuum gap, specifically the section between the membrane metallization and the antennas. The role of the antennas is to strongly focus the microwave electric fields into the vacuum gap. SiN membranes under such coupling to microwave resonators have demonstrated procedures such as sideband\cite{yuan2015large, noguchi2016ground} and feedback\cite{rej2025near} cooling, and quantum backaction evading measurements\cite{liu2022quantum}. 




The flipchip sits in a grounded aluminum sample holder with a coaxial cable and SMA connector serving as the input from a waveform generator, allowing DC and AC voltages to be applied between the gate electrode and the membrane. To facilitate the capacitive coupling between the membrane and the antenna chip structures, a 50\,nm thick, 0.5\,mm square aluminum pad is deposited on the membrane, before gluing the gold sphere, by evaporation through a physical mask. \figref{fig: device1}(b)\&(c) show schematics of the flipchip and \figref{fig: device1}(d) shows the chip mounted in the sample holder.

An equivalent circuit diagram for the electrostatic driving scheme is shown in \figref{fig: device1}(a), demonstrating the various capacitive couplings in the system. One can immediately conclude that the voltage across the gate-membrane capacitance should not be equal to the voltage output from the generator, and so both analytical calculations and circuit simulations are used to determine the various voltage divisions, modeling each component as a parallel plate capacitor. The model relies on voltage division pre-factors, where if the generator voltage is $V$, the voltage between the gate and membrane is $V_\mathrm{gm} = p_\mathrm{gm}V$ and the voltage between the membrane and antennas is $V_\mathrm{ma} = p_\mathrm{ma}V$. Here, the dimensionless factors $p_\mathrm{gm}$ and $p_\mathrm{ma}$ denote the voltage division in question. We denote the overlapping electrode area between the gate and membrane metallization as $S_\mathrm{gm}$ and the total overlapping membrane-antenna area as $S_\mathrm{ma}$, such that a total area is $S_\mathrm{tot} = S_\mathrm{gm} + S_\mathrm{ma}$. The electrostatic force on the membrane, derived from the electrostatic energy, is

\begin{equation}
\label{eq:Fd}
    F_\mathrm{d} = \frac{\epsilon_0 S_\mathrm{gm} p_\mathrm{gm}^2 V^2}{2 d^2} + \frac{\epsilon_0 S_\mathrm{ma} p_\mathrm{ma}^2 V^2}{2 d^2} = \frac{\epsilon_0 S_\mathrm{tot} p^2 V^2}{2 d^2},
\end{equation}

\noindent where $d$ is the gap size between the membrane and electrodes. An effective voltage division pre-factor of

\begin{equation}    \label{eq: pre-factor}
    p = \left( \frac{ p_\mathrm{gm}^2 S_\mathrm{gm} + p_\mathrm{ma}^2 S_\mathrm{ma} }{S_\mathrm{tot}} \right)^{1/2}
\end{equation}

\noindent has been introduced here to express the total electrostatic force.
Under a DC voltage, the electrostatic force in \eref{eq:Fd} induces a static displacement $x_\mathrm{s}$ of the membrane from equilibrium towards the electrodes. To model the frequency shift due to the voltage $V_\mathrm{DC}$, the membrane restoration spring force equal to $M \omega_\mathrm{m}^2 x_\mathrm{s}$ must be balanced with the electrostatic force, allowing one to calculate, for example, the effective mechanical resonance frequency, given by

\begin{equation}    \label{eq: omega_eff}
    \omega_\mathrm{eff} = \sqrt{ \omega_\mathrm{m}^2 - \frac{\epsilon_0 S_\mathrm{tot} p^2 V_\mathrm{DC}^2}{M(d - x_\mathrm{s})^3} }.
\end{equation}

\noindent 
The maximum frequency, which equals the intrinsic frequency, is obtained at $V_\mathrm{DC} = 0$. However, if there is any charge build up on the membrane, a static offset of the DC voltage, $V_0$, will cause the effective voltage to be $V_\mathrm{DC} - V_0$ and the maximum frequency will be shifted. This effect is observed in our experimental measurements.

The large amount of mass loading and nearby antenna chip, together with
the optomechanical implementation add complications when predicting the device behavior. To gain fundamental understanding of the mechanisms which influence the behavior, measurements were taken at room temperature as discussed in section \ref{sec: RT measurements}, which provide direct access to measurements of the membrane oscillation amplitude. Measurements at cryogenic temperatures are discussed in section \ref{sec: cryo}, allowing the microwave cavity to be utilized to readout the mechanical response.


\subsection{Microwave cavity}

In order to investigate the device in a dilution refrigerator, we use the microwave-optomechanical readout via a superconducting microwave cavity. Previous studies have demonstrated optomechanical systems based on this concept \cite{yuan2015large, yuan2015silicon, noguchi2016ground, liu2021gravitational, liu2022quantum, depellette2025amplitude}. The difference to traditional design is the introduction of DC (or, low-frequency) voltage. DC bias voltage application to superconducting resonators at large has been introduced in earlier work, both for planar resonators \cite{Simmonds2011DCbias}, as well as 3D designs \cite{Steele2017DCbias,Wallraff2018DCbias,Pate2018DCbias,Xia2024DCbias}. In contrast to purely electric field dipole coupling used with eg., superconducting qubits, our parametric coupling is very weak, and reaching an appreciable coupling in practice requires as strong as possible focusing of the electric fields inside the vacuum gap. However, the vacuum gap is the same place where the voltage needs to be applied in order to create the actuation via variable capacitance. Without a proper design, the voltage electrode acts as a strongly coupled external coupler which will drastically reduce the quality factor of the cavity down to far too low values.


\begin{figure}
    \centering
    \includegraphics[width=0.99\linewidth]{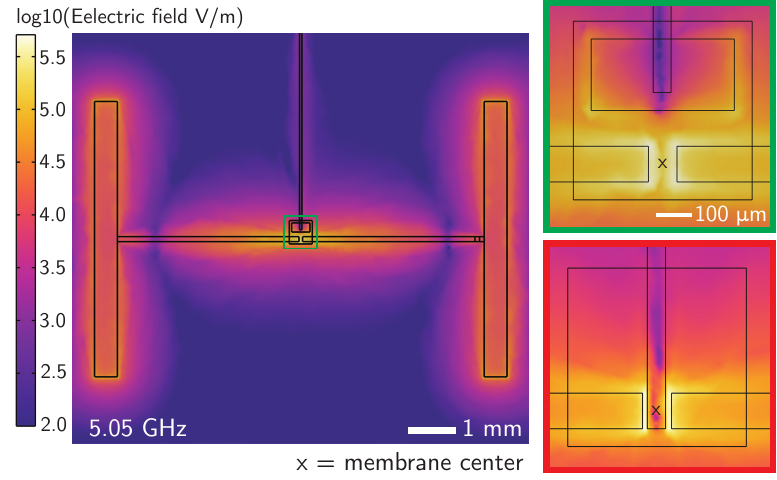}
    \caption{Simulating the electromagnetic mode inside the cavity using COMSOL. The antennas focus the electric field to produce a mode at 5.05\,GHz, with a particularly large field strength in the center. The membrane metallization produces a mechanically compliant capacitor with the antennas to couple the cavity mode to the membrane displacement. The green box shows a variant of the antenna geometry which produces a large field at the membrane center while still having the gate electrode close by. The red box shows a less optimized design which perturbs the field at the center of the membrane, reducing the coupling strength.}
    \label{fig: COMSOL}
\end{figure}

Finite element modeling is done in COMSOL to determine the properties of the microwave mode inside the cavity. The system begins with 16 mm x 12 mm x 3 mm aluminum cavity, with a fundamental electromagnetic mode of the empty cavity at 52\,GHz. The mode is strongly modified by introducing the chips and antennas, bringing the mode frequency to around 5\,GHz. Multiple electrode geometries have been simulated and fabricated to optimize a balance between the optomechanical coupling and the capability for electrostatic manipulation. \figref{fig: COMSOL} shows the electrode design used for the measurements discussed in section \ref{sec: cryo} and the simulated electric field inside the cavity. The gate electrode in this case is far enough away from the two antennas such that the electric field at the center of the membrane is not particularly modified by its presence, while keeping a significant overlap between the gate and membrane metallization. An alternative geometry is also simulated to show how the field is reduced when the gate overlaps with the membrane center, decreasing the optomechanical coupling.

With the gate electrode connected directly to a 50\,$\Omega$ transmission line, the simulated microwave mode has a quality factor of around $10^4$, which is already a reasonable value for the experimental applications in this study. One loss mechanism originates from the microwave field leaking to the gate transmission line, and so a low pass filter consisting of a 500\,nF capacitance to ground was implemented to reduce the leakage, while still allowing DC and low kilohertz AC voltage to pass to the gate. Including the filter in the simulations leads to an increase in the cavity mode quality factor by one order of magnitude.

\section{Room temperature measurements} \label{sec: RT measurements}


In order to mitigate the viscous air damping experienced by the membrane, the sample holder containing the flipchip is mounted inside a vacuum chamber, pumped by both a scroll pump and a turbo pump. Measurements with the turbo pump off are taken at a pressure of 0.1\,mbar, and at $9\times 10^{-4}$\,mbar with the turbo pump on. The chamber contains an SMA feed-through port which allows the sample holder to be connected to an external voltage generator. We use a laser Doppler vibrometer to measure the membrane oscillations through a window in the chamber. 



\subsection{Mechanical properties and frequency tuning}

\begin{figure}
    \centering
    \includegraphics[width=0.9\linewidth]{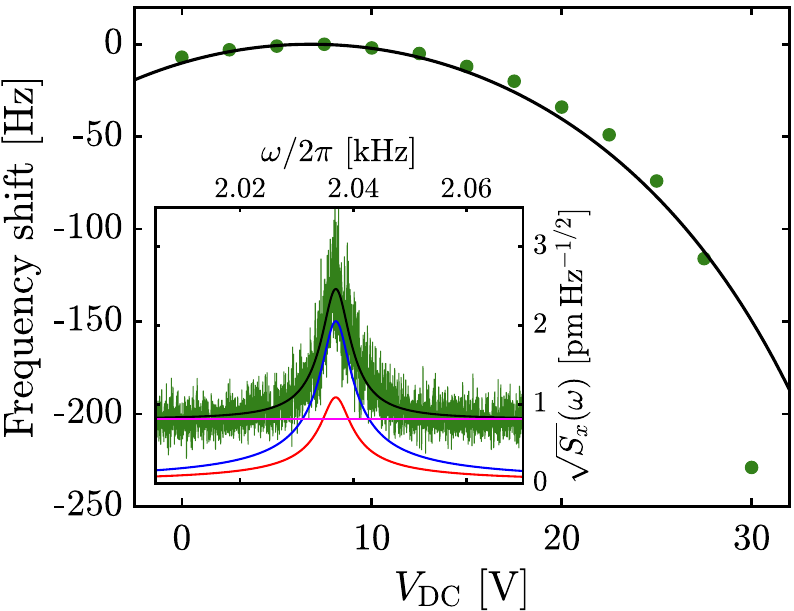}
    \caption{Tuning of the mechanical resonance frequency by applying a DC voltage between the membrane and gate electrode. The shift of frequency from its unbiased value is obtained by locating the mechanical resonance in the displacement spectrum, measured with a vibrometer. The black line is a theoretical fit determined by \eref{eq: omega_eff}. Inset: An example of the measured displacement spectral density of the membrane (green), for the fundamental mechanical mode. The effective temperature of the mode is calculated by the area under the curve, with the black line showing the corresponding theoretical spectrum for this temperature, given by \eref{eq: FDT} after taking the background noise into account. The magenta line shows the background noise due to measurement imprecision, the red and blue lines show the thermal and vibrational contributions, respectively, to the displacement spectrum.}
    \label{fig: DC tuning}
\end{figure}

Identification of the mechanical mode begins with a vibrometer measurement of the gold sphere position spectrum, initially with no voltages applied and with the turbo off. In this instance, the mechanical oscillations are driven by a combination of thermal excitations and vibrations in the system, with the resulting displacement spectral density $\sqrt{S_x}(\omega)$ being shown in the inset of figure \figref{fig: DC tuning}. With the expected resonance frequency of approximately 2\,kHz, and the spectrum being free of other vibrational modes close to this this frequency, the fundamental mechanical mode was easy to distinguish. A Lorentzian fit to the data gives a resonance frequency of $\omega_\mathrm{m}/2\pi = 2.04$\,kHz and a damping rate of $\Gamma_\mathrm{m}/2\pi = 5$\,Hz.

With the obtained mechanical parameters one can estimate the thermal and vibrational contributions to the overall displacement. Firstly, the y-offset due to measurement imprecision noise, shown with the magenta line, is removed from the spectrum. The area under the curve of the power spectral density, $S_x(\omega)$, directly gives the position variance $\left< x^2 \right>$. One can then calculate the effective temperature using the equipartition theorem

\begin{equation}
    T_\mathrm{eff} = \frac{M\omega_\mathrm{m}^2 \left< x^2 \right>}{k_\mathrm{B}},
\end{equation}

\noindent resulting in $T_\mathrm{eff} = 1334$\,K. This effective temperature has an associated Langevin force with a flat spectral density $S_f~=~2\Gamma_\mathrm{m}Mk_\mathrm{B}T_\mathrm{eff}$, such that the mechanical oscillator with susceptibility $\chi_\mathrm{m}(\omega)$ responds with a displacement spectrum

\begin{equation}    \label{eq: FDT}
    \sqrt{S_x}(\omega) = |\chi_\mathrm{m}(\omega)| \sqrt{S_f},
\end{equation}

\noindent which is shown with the black line in the inset of \figref{fig: DC tuning}, after re-adding the background noise. The contribution from the thermal environment is shown with the red line, calculated by replacing the temperature in \eref{eq: FDT} with $T_\mathrm{RT} = 293$\,K. The remaining contribution due to vibrations is assigned an effective temperature of $T_\mathrm{V} = T_\mathrm{eff} - T_\mathrm{RT} = 1041$\,K, with the blue line showing the spectrum calculated using \eref{eq: FDT} for this temperature. One may model the vibrational contribution as an inertial actuation of the membrane, caused by the environment having a displacement spectral density of $\sqrt{S_x^\mathrm{E}}(2\,\mathrm{kHz}) = 5$\,fm\,Hz$^{-1/2}$. 

To summarize, the mechanical mode is not in thermal equilibrium with the room temperature, largely due to vibrations of the scroll pump. When the turbo is switched on, the effective temperature is observed to increase by a factor of 2.5, which is expected from the increased vibrations. The mechanical resonance frequency also decreases to $\omega_\mathrm{m}/2\pi = 1.95$\,kHz due to the decrease in pressure. The significant actuation out of thermal equilibrium does not, however, cause problems for measuring the membrane displacement due to electrostatic actuation, which is orders of magnitude greater. 

The ability to measure the mechanical oscillations without additional driving
already allows for the observation of frequency tuning. A DC voltage is applied between the gate and the membrane, causing a static displacement in the equilibrium and decreasing the effective spring constant. Shown in \figref{fig: DC tuning}, the change in frequency is measured as a function of voltage by measuring the location of the mode which appears in the displacement spectrum. The theoretical prediction in \figref{fig: DC tuning} (black line) shows the result from \eref{eq: omega_eff} with $V_0$ as a fit parameter, giving an agreement with the majority of the data when using $V_0 = 6.8$\,V. The pre-factor used here is $p = 0.14$, obtained by measuring the membrane actuation which is discussed in section \ref{sec: El actuation}.


The ability to tune the resonance provides further evidence that the correct mode has been identified.
A maximum frequency shift of 230\,Hz is measured, confirming the potential to tune the source mass to the resonance of a similarly fabricated force detector. A large number of our devices have been previously measured with the frequencies typically varying by a few hundred hertz, and so the tunability observed here is reasonably satisfactory.

\subsection{Electrostatic actuation}    \label{sec: El actuation}

\begin{figure}
    \centering
    \includegraphics[width=0.9\linewidth]{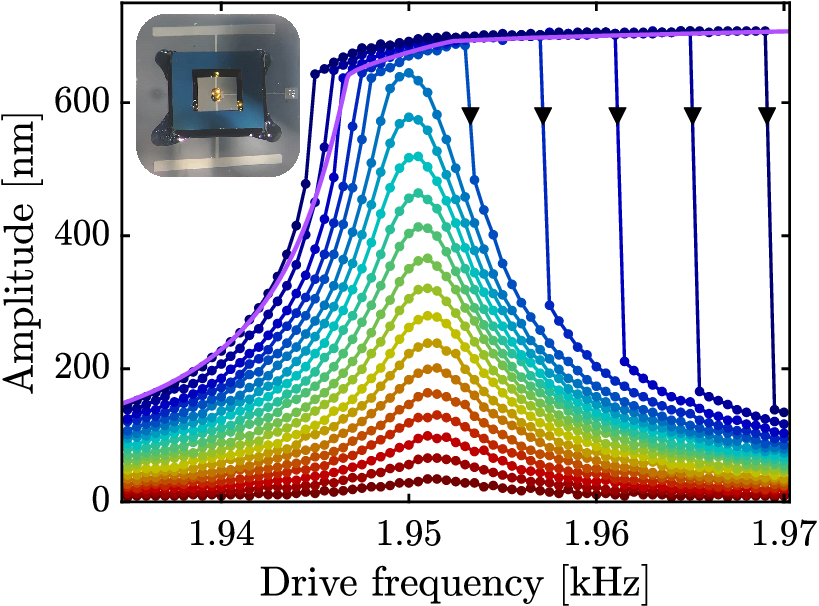}
    \caption{Electrostatically actuated flipchip membrane (pictured in the inset). Each colored dataset shows the measured membrane oscillation amplitude when an AC drive is swept in frequency through the mechanical resonance in the positive direction, from 0.25\,V (red) to 5\,V (blue). At large amplitudes the response shows a strong hardening and discontinuous jumps due to a repulsive contact force between the membrane and chip below. The purple line shows the simulated response in the presence of a contact force for the largest drive.}
    \label{fig: main actuation}
\end{figure}

After identifying the mechanical mode and confirming its response to an applied voltage, the response due to electrostatic driving is measured. A DC voltage with magnitude $V_\mathrm{DC} = 4\,$V is applied from the generator. An AC voltage is then used to drive the membrane, with \figref{fig: main actuation} showing the response measured with the vibrometer. Each colored dataset shows the measured gold sphere oscillation amplitude for a different AC voltage, from $V_\mathrm{AC}=0.25$\,V (red) to $V_\mathrm{AC}=5$\,V (blue). The drive frequency is swept in the positive direction through the mechanical resonance for each voltage, with the lower responses corresponding to a linear harmonic oscillator. Quantitative analysis of the linear responses relies on knowing the gap size $d$, which can be inferred from the larger amplitude, nonlinear, responses. Above an amplitude of 640\,nm, the measured response experiences the sudden onset of a strong hardening, with the gradient discontinuity suggesting the presence of a piecewise force which only engages above a certain amplitude. As the drive power is increased beyond this point the maximum amplitude continues to grow at a much smaller rate than in the linear regime, no longer being proportional to $V_\mathrm{AC}$, and eventually saturates at 708\,nm.

A mathematical model of the response in this strongly nonlinear regime relies on introducing a contact force in the equation of motion\cite{hertz1896miscellaneous}: physically, this is to say that the membrane experiences collisions when it is driven at an amplitude close to the gap size. The exact nature of the collisions is difficult to assess, but due to the nonlinearity engaging roughly 70\,nm before the saturation point, it is likely that compressible pieces of dust contribute to the overall force. We propose that the hard saturation limit is given by the gap size itself, with collisions between the membrane and antenna chip occurring at this amplitude. Thus, the value of $d = 708$\,nm is used for the analysis pertaining to this device which follows.

Theoretical and experimental studies of mechanical oscillators colliding with surfaces are prominent in the atomic force microscopy community\cite{lee2002nonlinear,holscher2006theory,kalafut2021cantilever,mendova2024size}, with repulsive contact also having wider applications for micro-electromechanical systems\cite{miller2021amplitude}. In line with previous models, a piecewise force is introduced in the equation of motion, where for an amplitude $x$ the additional force on the membrane is

\begin{equation}    \label{eq: CF}
    F_\mathrm{CF} = -\omega_\mathrm{CF}^2M (x - g) H(x - g),
\end{equation}

\noindent where $H(x - g)$ is the Heaviside function such that the force only engages above an amplitude of $g$. The strength of this linear hardening term is characterized by a stiffness $\omega_\mathrm{CF}$, often calculated by knowing the elastic moduli and geometry of the colliding surfaces. If the contact surface contains an asperity of radius $R$, in this case given by dust or surface roughness, the stiffness parameter is

\begin{equation}    \label{eq: omegaCF}
    \omega_\mathrm{CF} = \sqrt{ \frac{2E^*R}{M} },
\end{equation}

\noindent where

\begin{equation}
    \frac{1}{E^*} = \frac{1 - \nu_1^2}{E_1} + \frac{1 - \nu_2^2}{E_2},
\end{equation}

\noindent where $E_i$ are the Young's moduli and $\nu_i$ are the Poisson ratios of the two contacting materials\cite{schwarz2003generalized}. With the unknown nature and location of the dust between the membrane and antenna chip, along with the possibility of having multiple contact points, such calculations are challenging for the system in question. Additional components in the contact force may also contribute to an increase in both viscous and non-linear damping\cite{miller2021amplitude}, possibly originating from squeeze-film damping due to air in the gap, however we choose not to add extra free parameters to the model due to the lack of rigorous physical justification.

A simulation using MATCONT\cite{dhooge2008new} is implemented to predict the response when the force in \eref{eq: CF} is added to the linear harmonic oscillator equation of motion. The procedure relies on numerically calculating the time domain solution for a single drive frequency close to resonance and then using numerical continuation to obtain the amplitudes at adjacent frequencies. By observing the five datasets in the nonlinear regime it appears that there are two distinct effects: at 640\,nm the contact force engages with a moderate hardening, while the amplitude response has a visibly positive slope as the drive frequency is swept; for the largest drive forces where the amplitude is closer to the gap size, however, the gradient of the response is smaller which suggests a stronger hardening term at these amplitudes. To account for both of these observations, the simulation uses a summation of two independent linear contact forces: the first engaging at an amplitude of $g_1 = 640$\,nm with stiffness $\omega_\mathrm{CF,1} = 8$\,kHz, and the second with a threshold amplitude $g_2 = 692$\,nm and stiffness $\omega_\mathrm{CF,2} = 38$\,kHz. The value of $\omega_\mathrm{CF,1}$ is chosen phenomenologically to match the gradient of the measured response for the $V_\mathrm{AC} = 4$\,V dataset, which is the first response in the nonlinear regime. Physically, this could be interpreted as one or several dust particles with radii $\sim 30$\,nm, providing the onset of repulsive contact. The second force parameters are chosen to match the gradient of the response close to the saturation limit, likely caused by additional collisions with smaller dust particles or surface roughness on the antenna.

The simulated response corresponding to the largest drive force used in the experiment is shown with the purple line in \figref{fig: main actuation}. The simulation includes the gradient discontinuity and strong hardening, however it does not capture the discontinuous jump-down in amplitude. Extra dissipation in the system can cause shifts in the bifurication frequency, destabilizing the high-amplitude branch earlier than predicted, and so it is possible that the previously mentioned squeeze film damping contributes to the overall contact force. The additional parameters required to model this are challenging to estimate with the current data, however the similarity between the data and theory suggests that repulsive contact is a likely explanation for the nonlinear response. 

With the gap size $d = 708$\,nm being estimated from the saturation point, the linear responses can now be analyzed in more detail. Fits to these data, given by the solution to the harmonic oscillator equation of motion, allow the fit parameters $\omega_\mathrm{m}$, $\Gamma_\mathrm{m}$, and the total force $F_\mathrm{d}$ to be extracted as functions of $V_\mathrm{AC}$. The gradient of a linear fit between $F_\mathrm{d}$ and $V_\mathrm{AC}$ is a function of both $V_0$ and $p$.
We find that the same parameter combination $(V_0, p)$  simultaneously satisfies both the data obtained from frequency tuning and electrostatic actuation.

Knowledge of the gap size also allows the capacitance between each electrode to be calculated, with the areas $S_\mathrm{gm}~=~(75 \,$$\mu$m)$^2$ and $S_\mathrm{ma}=(283 \,$$\mu$m)$^2$. This allows for a theoretical prediction of $(p_\mathrm{gm}, p_\mathrm{ma}) = (0.89, 0.062)$ to be found through circuit simulation, leading to a predicted $p = 0.24$ using \eref{eq: pre-factor}. A discrepancy with the simulated and measured values of $p$ may be explained by the static offset changing over time. The measured frequency shift at $V_\mathrm{DC} = 30$\,V does not follow the theoretical prediction, suggesting that discharges may occur over the course of the experiment. The possibility of squeeze-film damping contributing to the amplitude saturation could also cause the gap size to be underestimated.

Despite some inconsistencies, the important physical aspects determining the driven response have been uncovered, notably demonstrating that the membrane is capable of being driven close to the gap size and shows the presence of a repulsive contact force. The source mass is capable of both frequency tuning by hundreds of hertz and coherent driving up to hundreds of nanometers, thus meeting our criteria for a controllable gravitational source. 





\subsection{Elevated membrane actuation}



Through the nonlinear dynamics, we have observed the collisions with the antenna chip, which naturally limit the actuation amplitude. The following section discusses an alternative device design, taking a mass loaded membrane in the absence of an antenna chip. We refer to this as an elevated membrane, designed to provide a confirmation that the previous results originate from a contact force and not an intrinsic mechanism of mass loaded membranes. This is also used to demonstrate how large the amplitudes can reach when the primary limitation is removed. The membrane chip frame is glued at each corner to four 500\,$\mu$m thick support pillars which are mounted on a silicon carrier chip. The suspension allows the membrane to oscillate at much greater amplitudes than the previously restricted device. The entire elevated membrane device is placed on a piezoelectric disc which is suspended in a plastic holder for inertial actuation. 
When the membrane is driven on resonance the piezo amplitude is resonantly enhanced by the mechanical quality factor.

\begin{figure}
    \centering
    \includegraphics[width=0.9\linewidth]{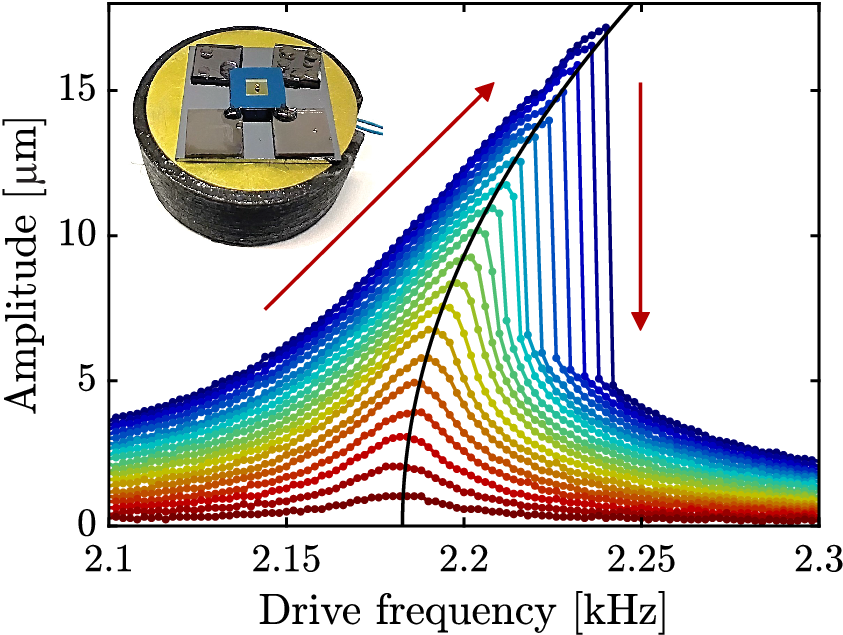}
    \caption{Inertial actuation of a membrane suspended from four support pillars, placed on a piezoelectric disc (pictured in the inset). Each colored dataset shows the measured membrane oscillation amplitude when an AC drive to the piezoelectric is swept in frequency through the mechanical resonance in the positive direction, from 0.2\,V (red) to 4\,V (blue). The response is characteristic of a Duffing oscillator, with the black line showing the fitted backbone given by \eref{eq: Duffing backbone}.}
    \label{fig: Duffing measurement}
\end{figure}

Shown in \figref{fig: Duffing measurement}, the membrane amplitude is measured for a range of drive voltages to the piezo, from 0.2\,V (red) to 4\,V (blue). The drive frequency is swept in the positive direction through the mechanical resonance for each voltage, with the response becoming increasingly nonlinear as the amplitude increases. Above an amplitude of around 10\,$\mu$m the response experiences discontinuities characteristic of a Duffing oscillator, and drive frequency sweeps in the negative direction (not shown) confirm that the response is hysteretic. The response of a Duffing oscillator as a function of drive power features a backbone which describes how the frequency at which the amplitude maxima occur changes as those amplitudes increase. Geometric nonlinearity causes a hardening of the spring constant as displacements get larger, where if the maximum of each response is the set of amplitudes $a_\mathrm{max}$, their corresponding frequencies are \cite{nayfeh2024nonlinear,kozinsky2006tuning}

\begin{equation}    \label{eq: Duffing backbone}
    \omega_\mathrm{max} = \omega_\mathrm{m} \left( 1 + \frac{a_\mathrm{max}^2}{\sqrt{3}Qa_\mathrm{c}^2} \right),
\end{equation}

\noindent where $\omega_\mathrm{m}$ is the mechanical resonance frequency in the linear (low amplitude) regime and $Q$ is the quality factor. The critical amplitude $a_\mathrm{c}$ quantifies the strength of the nonlinearity and is the amplitude above which the response becomes discontinuous and hysteretic. 

The backbone curve in \figref{fig: Duffing measurement} (black line), given by \eref{eq: Duffing backbone}, is a fit to the maximum amplitude of each response and the frequency at which they occur, with fit parameters: $\omega_\mathrm{m}/2\pi~=~2.18$\,kHz, $Q = 63$, and $a_\mathrm{c} = 10$\,$\mu$m. This is consistent with geometric hardening of the membrane from Euler-Bernoulli theory, which predicts that the critical amplitude should be on the order of several microns. This is much greater than the several nanometer critical amplitudes of unloaded SiN membranes previously observed\cite{zwickl2008high}, owing to the large mass loading and low quality factor. 

The clear difference between the measured response of the vacuum-gap and elevated membranes, notably the several microns of actuation and the agreement with the theory of Duffing oscillators in the elevated membrane case, demonstrates that the mass loaded membrane does not have an intrinsic mechanism which causes strong, contact force style, hardening and amplitude saturation. This is an important demonstration showing that the devices are capable of higher oscillation amplitudes and the only limit on amplitude during electrostatic actuation is given by the gap size, which is easily adjustable in the fabrication.
Cryogenic measurements are discussed next, which enable significantly higher mechanical quality factors than at room temperature. This allows for additional resonant enhancement from the driving, which can enable high amplitudes with an increased gap size while still using reasonable voltages.  


\section{Cryogenic measurements}    \label{sec: cryo}



We now demonstrate the microwave-optomechanical implementation of the system, which requires the aluminum films to be superconducting, and thus sub-Kelvin temperatures. The measurement goals are largely similar to those previously discussed, namely identification of the mechanical mode and confirming the frequency tuning and actuation capabilities.

\subsection{Experimental setup}


The flipchip sample holder used in the room temperature measurements already provides a suitable 3D cavity when a lid is placed to enclose the system. The extended central pin bonded to the gate electrode remains in place, while an additional, perpendicular, SMA pin is used to provide an external coupling of the cavity to a transmission line. Coaxial lines connect both the sample inputs to the room temperature equipment, with the mentioned capacitive low pass filter on the gate electrode line preventing cavity losses through that channel. The measurements are carried out in a dry dilution refrigerator at a temperature of 10 mK.
Multiple flipchip devices have undergone actuation measurements in this setup, and due to the unpredictable nature of fabricated samples, not all of them are detectable in the cooldown. This includes the original device measured at room temperature, and so this section instead discusses a similar device with the same fabrication recipe.

The cavity resonance is measured with a VNA (see \figref{fig: actuation cryo}) at a frequency of $\omega_\mathrm{c}/2\pi = 5.05$\,GHz, with internal and external energy decay rates $\kappa_\mathrm{i}/2\pi= 2.8$\,MHz and $\kappa_\mathrm{e}/2\pi =500$\,kHz. The external coupling is determined by the length of the SMA pin which extends into the cavity, while the internal losses are determined by a combination of material losses inside the cavity and microwaves leaking to the gate port. The two internal contributions are indistinguishable in the measurement and are larger than the simulated values from COMSOL, however the simulations do not take into account material losses. Larger than expected internal losses are a commonly observed problem when placing extra components inside 3D cavities, which would otherwise have a higher quality factors when empty. In this particular study, the cavity losses do not disrupt the ability to conduct the relevant measurements. Readout of the mechanical resonance is achieved by measuring the cavity output spectrum with a DAQ; by reflecting a weak microwave tone from the cavity, the mechanical oscillations imprint sidebands on the spectrum due to the optomechanical coupling. A Lorentzian fit to the measured power gives a mechanical resonance frequency of $f_\mathrm{m} = 1.66$\,kHz and a quality factor of $10^5$.


\subsection{Frequency tuning}

\begin{figure}
    \centering
    \includegraphics[width=0.8\linewidth]{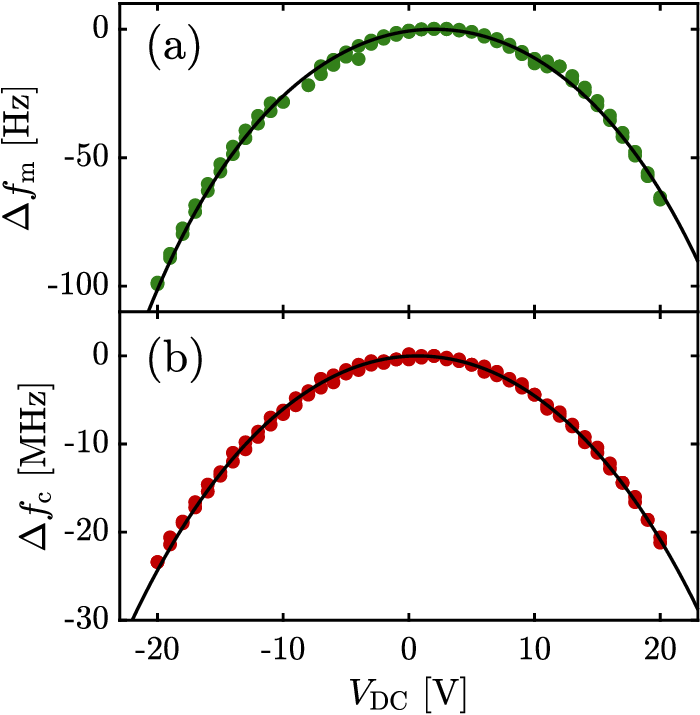}
    \caption{Frequency tuning of the (a) mechanical and (b) cavity resonance frequencies. A DC voltage is applied between the gate electrode and membrane, and the mechanical resonance is detected using a spectrum analyzer to measure the output spectrum sideband. The cavity resonance is detected with a VNA. The black lines are theoretical fits based on \eref{eq: omega_eff} and \eref{eq: fc vs x}.}
    \label{fig: DC cryo}
\end{figure}

Similar to the measurement in \figref{fig: DC tuning}, a DC voltage is applied between the gate and membrane, shifting the equilibrium position and changing the resonance frequency, shown in \figref{fig: DC cryo}(a). The change in frequency $\Delta f_\mathrm{m}$ from its unbiased value follows from \eref{eq: omega_eff}, with the theoretical prediction shown with the black line. Here, we use the same pre-factor of $p = 0.14$ as found in the room-temperature measurements, while the unbiased gap size $d = 915$\,nm remains as a fitting parameter. The fitted gap size here is reasonable based on the interference pattern visible beneath the membrane and is within the typical range of our fabricated devices. One can achieve very similar theoretical results for a range of $(p,d)$ combinations, and with the laser vibrometer not available for cryogenic measurements, an independent verification of either parameter is challenging, hence the fixing of $p$ based on previous data. Furthermore, a room temperature measurement of the gap size would not be applicable at cryogenic temperatures due to thermal contractions influencing the value of $d$. 

As the DC voltage displaces the membrane, the gap size decreases and therefore the membrane-antenna capacitance increases. The cavity itself is modeled as a lumped element $LC$ resonator, and due to the changing system capacitance, the cavity resonance frequency has a corresponding shift $\Delta f_\mathrm{c}$ as a function of the applied voltage, shown in \figref{fig: DC cryo}(b). The theoretical prediction (black line) follows the most basic principle of optomechanics: a displacement $x_\mathrm{s}$ causes a proportional shift in cavity frequency

\begin{equation}    \label{eq: fc vs x}
    \Delta f_\mathrm{c} = - G x_\mathrm{s},
\end{equation}

\noindent where $G$ is the strength of the optomechanical coupling. The static displacement as a function of $V_\mathrm{DC}$ is already numerically calculated from the fit to the shift in mechanical frequency, and so the theoretical prediction follows by taking $G = 0.55$\,MHz\,nm$^{-1}$ as a fitting parameter. The order of magnitude of $G$ agrees with finite element simulations and has been observed with a number of our similarly fabricated devices. While electrostatic manipulation of the mechanical oscillations is the primary goal of this device, the additional consequence of cavity tuning is beneficial, having applications towards cryogenic filtering used to decrease generator noise. 


\subsection{Actuation}

\begin{figure}
    \centering
    \includegraphics[width=0.8\linewidth]{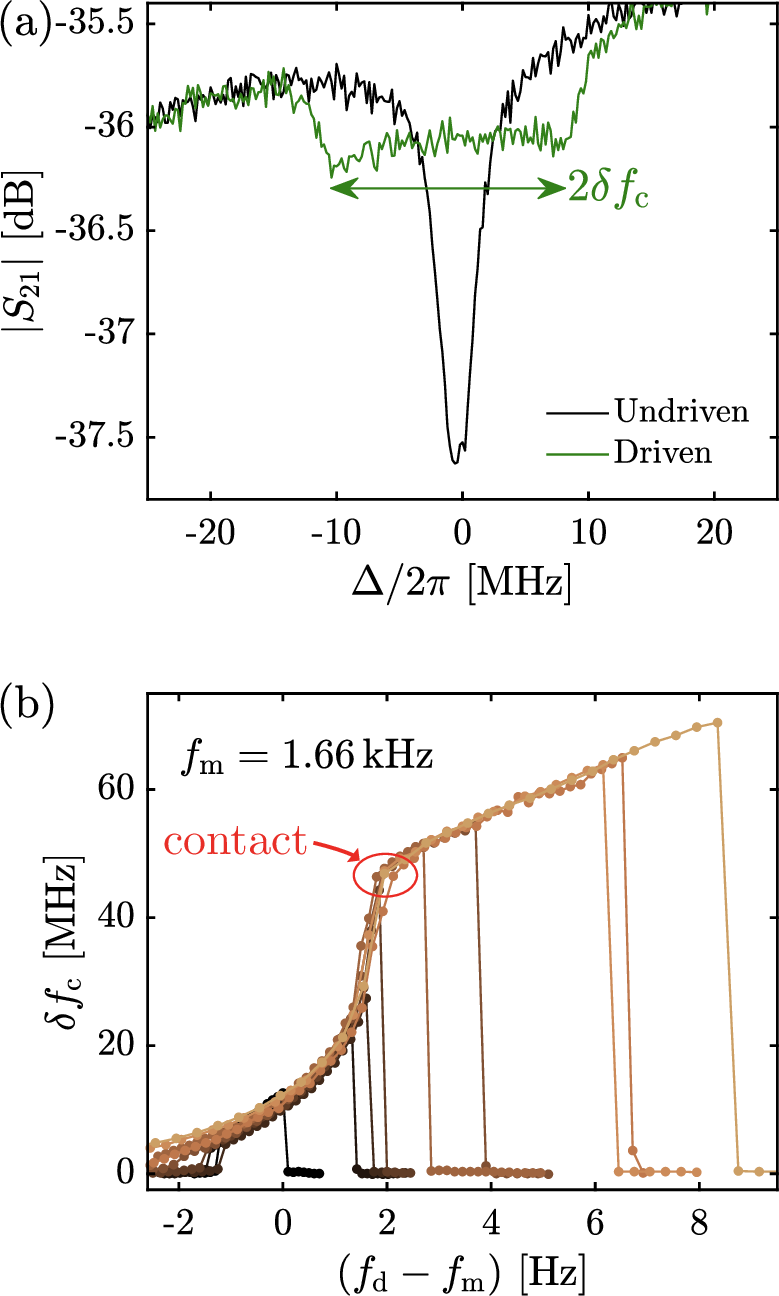}
    \caption{Electrostatic actuation of the flipchip membrane inside a 3D cavity. (a) The cavity resonance is measured with a VNA while the membrane is in its natural oscillation state (undriven) and electrostatically actuated state (driven). The driven oscillations cause the cavity resonance to be modulated, with a resulting profile width of $2\delta f_\mathrm{c}$. (b) The modulated widths of each cavity resonance are used to characterize the membrane response when driven with an AC voltage with frequency $f_\mathrm{d}$ swept in the positive direction, from 15\,mV (dark brown) to 200\,mV (light brown).}
    \label{fig: actuation cryo}
\end{figure}

Following the same procedure as the room temperature experiment, a 4\,V DC voltage is applied while the AC drive frequency is swept through the mechanical resonance to actuate the membrane position. We denote the amplitude of the resulting cavity frequency modulation by $\delta f_c$, given in the same form as \eref{eq: fc vs x} for the static shift.
 Since $\delta f_c \gg \kappa$, the optomechanical readout is fully nonlinear. The motion of the membrane is therefore inferred from the modulated cavity resonance, which oscillates in frequency due to the optomechanical coupling. Figure \figref{fig: actuation cryo}(a) shows a measured cavity resonance without electrostatic driving (black) and an example when the membrane is driven on resonance (green). The VNA essentially measures a time average of the resonance moving back and forth in frequency, 
 producing a characteristic `w' shape. The local minima in reflected power are separated by a frequency $2\delta f_\mathrm{c}$.
 A reliable conversion between $\delta f_\mathrm{c}$ and the membrane amplitude was not found during the measurements, largely due to the unknown voltage at the device level and uncertainty in gap size preventing consistency checks for the calibration. The observation of a modulated cavity resonance, however, does imply that the membrane is successfully driven by the applied voltage.

To extract the modulation amplitudes, the corresponding cavity resonance traces are measured for each pair $(V_\mathrm{AC}, f_\mathrm{d})$ of AC voltage and drive frequency. The measured $\delta f_\mathrm{c}$ for frequency sweeps in the positive direction are shown in \figref{fig: actuation cryo}(b) for a range of AC voltages, from 15\,mV (dark brown) to 200\,mV (light brown). The mechanical resonance frequency was observed to drift over the timescale of the experiment by several Hz, and so each trace has been manually corrected to display the drive detuning $f_\mathrm{d} - f_\mathrm{m}$ as opposed to the absolute drive frequency $f_\mathrm{d}$. 

At low AC voltages the response displays a complex combination of hardening and softening, with the multiple effects of electrostatics, geometric hardening, higher order optomechanical coupling, and drifting mechanical frequency all likely contributing to the nonlinearity even at small amplitudes. For example, the traces corresponding to the four lowest AC voltages each show discontinuities in amplitude above and below the mechanical resonance frequency, suggesting an `s' shaped response reminiscent of previously observed NEMS devices\cite{samanta2018tuning}. Frequency sweeps in the negative direction and at lower AC voltages (not shown) confirm that the response is hysteretic and experiences both hardening and softening nonlinearities. While the response has many intricacies, the primary goal of having a coherently driven source mass as part of an optomechanical system has been demonstrated here.

At $\delta f_\mathrm{c} = 46$\,MHz, all traces in the large $V_\mathrm{AC}$ regime show a discontinuity in the gradient of the measured response. As established in the room temperature measurements, this is indicative of a piecewise force which only engages above this amplitude, and so we interpret the observed response as a repulsive contact force. If one takes only the linear optomechanical coupling term, such that a membrane amplitude of $\delta x$ causes a modulated cavity frequency of $\delta f_\mathrm{c} = G \delta x$, this suggests that the membrane is oscillating with an amplitude of approximately 80\,nm. With the estimated gap size of $d = 915$\,nm, it may be concluded that at least one large piece of dust lies below the membrane and impacts long before the membrane comes close to the antenna. With the measured modulation amplitude continuing to increase to 70\,MHz and the absence of saturation, it is clear that the repulsive contact is not caused by a hard surface, such as the antenna itself, and instead should be attributed to a compressible asperity. Alternatively, the dust particle(s) may lie towards the edges of the membrane, allowing the center of the membrane to continue to be driven further, regardless of their compressibility.

To provide a confirmation that the contact force style nonlinearity is not a feature of the electrostatic driving force, measurements of the same device, in the same conditions, were conducted with an alternative actuation method. Similar to the elevated membrane case, a piezoelectric shaker was placed in the cryostat near to the 3D cavity sample holder in order to inertially actuate the membrane by vibrations. Sweeps in both drive frequency and power revealed the same discontinuity in response gradient at $\delta f_\mathrm{c} = 46$\,MHz, with the same moderately increasing amplitude beyond this point as was observed with the electrostatic actuation.

Overall, we interpret the results as a successful demonstration that the microwave optomechanical implementation does not introduce fundamental limitations in either the frequency tuning or actuation capability. No hard saturation limit on the actuation was reached, suggesting future improvements in the fabrication are required to remove dust more thoroughly, such that oscillations closer to the gap size can be reached.


\section{Conclusion}

We have demonstrated a mass loaded SiN membrane device which combines electrostatic actuation, frequency tuning, and microwave optomechanical implementation, which is compatible with the proposal by Liu \textit{et al.}\cite{liu2021gravitational} to measure the gravitational force between two 1\,mg masses. At room temperature, tunability of the mechanical resonance frequency by hundreds of hertz is demonstrated, providing the ability to match the resonance of a detector in future experiments. Furthermore, 700\,nm of coherently driven oscillation is observed, with repulsive contact between the membrane and close-by antenna chip being the limiting mechanism preventing larger amplitudes. This can be addressed in the future by modifying the fabrication procedure to allow for larger gap sizes, if required in future applications. Further measurements involving a suspended membrane were used to confirm that only the expected geometric hardening contributes to nonlinearity in the absence of a contacting surface, with no indication of amplitude saturation up to several micrometers of oscillation.

Cryogenic implementation allows the cavity and mechanical resonances to be measured using standard microwave techniques, with the antennas providing an optomechanical coupling between the membrane displacement and cavity resonance frequency. Tuning both of these with a DC voltage bias yields comparable success to the room temperature measurements, with 100\,Hz of mechanical frequency tuning and over 20\,MHz of cavity frequency tuning. During electrostatic actuation of the membrane, the modulated cavity was used to characterize the driven response, with a contact force style nonlinearity being observed and estimated amplitudes on the order of 100\,nm. With dust beneath the membrane being the predicted cause of repulsive contact, this appears to not be a fundamental limitation of the device design. Overall, the introduction of the microwave optomechanical setup does not appear to compromise the main purpose of the device, namely frequency tuning and actuation. Proven optomechanical techniques for quantum state preparation suggest future opportunities to manufacture a system sharing a measurable gravitational field and macroscopic quantum phenomena.

\begin{acknowledgments}
We would like to thank Longhao Wu for photography. We acknowledge the facilities and technical support of Otaniemi research infrastructure for Micro and Nanotechnologies (OtaNano). This work was supported by the Research Council of Finland (contract 352189), and by the European Research Council (contract 101019712). The work was performed as part of the Research Council of Finland Centre of Excellence program (contracts 352932, and 336810). We acknowledge funding from the QuantERA II Programme (contract 13352189).
\end{acknowledgments}

\section*{Author Declarations}

\subsection*{Conflict of Interest}

The authors have no conflicts to disclose.

\subsection*{Author Contributions}

\textbf{Joe~Depellette:} Conceptualization (equal); Data curation (equal); Formal analysis (lead); Investigation (equal); Resources (equal); Software (lead); Validation (equal); Visualization (lead); Writing - original draft (lead); Writing - review \& editing (lead). \textbf{Ewa~Rej:} Conceptualization (equal); Data curation (equal); Formal analysis (supporting); Investigation (equal); Resources (equal); Software (supporting); Validation (equal); Writing - review \& editing (supporting). \textbf{Richa~Cutting:} Investigation (supporting); Resources (equal). \textbf{Mika~A.~Sillanpää:} Conceptualization (equal); Funding acquisition (lead); Supervision (lead); Validation (equal); Writing - review \& editing (supporting).

\section*{Data Availability}

The data that support the findings of this study are available from the corresponding author upon reasonable request.

\section*{References}

\bibliography{References}

@article{Xia2024DCbias,
  title = {Gate-compatible circuit quantum electrodynamics in a three-dimensional cavity architecture},
  author = {Xia, Zezhou and Huo, Jierong and Li, Zonglin and Ying, Jianghua and Liu, Yulong and Tang, Xin-Yi and Wang, Yuqing and Chen, Mo and Pan, Dong and Zhang, Shan and Liu, Qichun and Li, Tiefu and Li, Lin and He, Ke and Zhao, Jianhua and Shang, Runan and Zhang, Hao},
  journal = {Phys. Rev. Appl.},
  volume = {21},
  pages = {034031},
  year = {2024},
  doi = {10.1103/PhysRevApplied.21.034031},
  url = {https://link.aps.org/doi/10.1103/PhysRevApplied.21.034031}
}

@article{Wallraff2018DCbias,
doi = {10.1088/2058-9565/aad362},
url = {https://doi.org/10.1088/2058-9565/aad362},
year = {2018},
volume = {3},
pages = {045007},
author = {Stammeier, M and Garcia, S and Wallraff, A},
title = {{Applying electric and magnetic field bias in a 3D superconducting waveguide cavity with high quality factor}},
journal = {Quantum Sci. Technol.},
}

@article{Pate2018DCbias,
    author = {Pate, Jacob M. and Martinez, Luis A. and Thompson, Johnathon J. and Chiao, Raymond Y. and Sharping, Jay E.},
    title = {{Electrostatic tuning of mechanical and microwave resonances in 3D superconducting radio frequency cavities}},
    journal = {AIP Adv.},
    volume = {8},
    pages = {115223},
    year = {2018},
    doi = {10.1063/1.5055887},
    url = {https://doi.org/10.1063/1.5055887},
}

@article{Steele2017DCbias,
    author = {Cohen, Martijn A. and Yuan, Mingyun and de Jong, Bas W. A. and Beukers, Ewout and Bosman, Sal J. and Steele, Gary A.},
    title = {{A split-cavity design for the incorporation of a DC bias in a 3D microwave cavity}},
    journal = {Appl. Phys. Lett.},
    volume = {110},
    pages = {172601},
    year = {2017},
    doi = {10.1063/1.4981884},
    url = {https://doi.org/10.1063/1.4981884},
}

@article{Simmonds2011DCbias,
    author = {Chen, Fei and Sirois, A. J. and Simmonds, R. W. and Rimberg, A. J.},
    title = {{Introduction of a dc bias into a high-Q superconducting microwave cavity}},
    journal = {Appl. Phys. Lett.},
    volume = {98},
    pages = {132509},
    year = {2011},
    doi = {10.1063/1.3573824},
    url = {https://doi.org/10.1063/1.3573824},
}

@article{liu2021gravitational,
  title = {Gravitational Forces Between Nonclassical Mechanical Oscillators},
  author = {Liu, Yulong and Mummery, Jay and Zhou, Jingwei and Sillanp\"a\"a, Mika A.},
  journal = {Phys. Rev. Appl.},
  volume = {15},
  issue = {3},
  pages = {034004},
  numpages = {22},
  year = {2021},
  month = {Mar},
  publisher = {American Physical Society},
  doi = {10.1103/PhysRevApplied.15.034004},
  url = {https://link.aps.org/doi/10.1103/PhysRevApplied.15.034004}
}

@article{kozinsky2006tuning,
  title={Tuning nonlinearity, dynamic range, and frequency of nanomechanical resonators},
  author={Kozinsky, I and Postma, H W Ch and Bargatin, I and Roukes, M L},
  journal={Appl. Phys. Lett.},
  volume={88},
  issue={25},
  pages={253101},
  number={25},
  year={2006},
  publisher={AIP Publishing},
  doi={https://doi.org/10.1063/1.2209211},
  url={https://pubs.aip.org/aip/apl/article/88/25/253101/330868/Tuning-nonlinearity-dynamic-range-and-frequency-of}
}

@book{nayfeh2024nonlinear,
  title={Nonlinear Oscillations},
  author={Nayfeh, Ali H and Mook, Dean T},
  year={1995},
  publisher={John Wiley \& Sons}
}

@article{zwickl2008high,
  title={High quality mechanical and optical properties of commercial silicon nitride membranes},
  author={Zwickl, B M and Shanks, W E and Jayich, A M and Yang, C and Bleszynski Jayich, A C and Thompson, J D and Harris, J G E},
  journal={Appl. Phys. Lett.},
  volume={92},
  pages={103125},
  number={10},
  year={2008},
  publisher={AIP Publishing},
  doi={https://doi.org/10.1063/1.2884191},
  url={https://pubs.aip.org/aip/apl/article/92/10/103125/145524/High-quality-mechanical-and-optical-properties-of}
}

@book{hertz1896miscellaneous,
  title={Miscellaneous papers},
  author={Hertz, Heinrich},
  year={1896},
  publisher={Macmillan}
}

@article{lee2002nonlinear,
  title={Nonlinear dynamics of microcantilevers in tapping mode atomic force microscopy: A comparison between theory and experiment},
  author={Lee, S I and Howell, S W and Raman, A and Reifenberger, R},
  journal={Phys. Rev. B.},
  volume={66},
  number={11},
  pages={115409},
  year={2002},
  publisher={APS},
  doi={https://doi.org/10.1103/PhysRevB.66.115409},
  url={https://journals.aps.org/prb/abstract/10.1103/PhysRevB.66.115409}
}

@article{kalafut2021cantilever,
  title={{Cantilever signature of tip detachment during contact resonance AFM}},
  author={Kalafut, Devin and Wagner, Ryan and Cadena, Maria Jose and Bajaj, Anil and Raman, Arvind},
  journal={Beilstein J. Nanotechnol.},
  volume={12},
  number={1},
  pages={1286--1296},
  year={2021},
  publisher={Beilstein-Institut},
  doi={https://doi.org/10.3762/bjnano.12.96},
  url={https://www.beilstein-journals.org/bjnano/articles/12/96}
}

@article{mendova2024size,
  title={{Size matters: rethinking Hertz model interpretation for cell mechanics using AFM}},
  author={Mendov{\'a}, Katar{\'\i}na and Ot{\'a}hal, Martin and Drab, Mitja and Daniel, Matej},
  journal={Int. J. Mol. Sci.},
  volume={25},
  number={13},
  pages={7186},
  year={2024},
  publisher={MDPI},
  doi={https://doi.org/10.3390/ijms25137186},
  url={https://www.mdpi.com/1422-0067/25/13/7186}
}

@article{holscher2006theory,
  title={{Theory of Q-controlled dynamic force microscopy in air}},
  author={H{\"o}lscher, H and Ebeling, D and Schwarz, U D},
  journal={J. Appl. Phys.},
  volume={99},
  number={8},
  pages={084311},
  year={2006},
  publisher={AIP Publishing},
  doi={https://doi.org/10.1063/1.2190070},
  url={https://pubs.aip.org/aip/jap/article/99/8/084311/292896/Theory-of-Q-Controlled-dynamic-force-microscopy-in}
}

@article{miller2021amplitude,
  title={Amplitude stabilization of micromechanical oscillators using engineered nonlinearity},
  author={Miller, James M L and Gomez-Franco, Ariosto and Shin, Dongsuk D and Kwon, Hyun-Keun and Kenny, Thomas W},
  journal={Phys. Rev. Res.},
  volume={3},
  number={3},
  pages={033268},
  year={2021},
  publisher={APS},
  doi={https://doi.org/10.1103/PhysRevResearch.3.033268},
  url={https://journals.aps.org/prresearch/abstract/10.1103/PhysRevResearch.3.033268}
}

@article{dhooge2008new,
  title={{New features of the software MatCont for bifurcation analysis of dynamical systems}},
  author={Dhooge, Annick and Govaerts, Willy and Kuznetsov, Yu A and Meijer, H Ga{\'e}tan Ellart and Sautois, Bart},
  journal={Math. Comput. Model. Dyn. Syst.},
  volume={14},
  number={2},
  pages={147--175},
  year={2008},
  publisher={Taylor \& Francis},
  doi={https://doi.org/10.1080/13873950701742754},
  url={https://www.tandfonline.com/doi/full/10.1080/13873950701742754#d1e425}
}

@article{schwarz2003generalized,
  title={A generalized analytical model for the elastic deformation of an adhesive contact between a sphere and a flat surface},
  author={Schwarz, Udo D},
  journal={J. Colloid Interface Sci.},
  volume={261},
  number={1},
  pages={99--106},
  year={2003},
  publisher={Elsevier},
  doi={https://doi.org/10.1016/S0021-9797(03)00049-3},
  url={https://www.sciencedirect.com/science/article/pii/S0021979703000493?via%3Dihub}
}

@article{samanta2018tuning,
  title={Tuning of geometric nonlinearity in ultrathin nanoelectromechanical systems},
  author={Samanta, Chandan and Arora, Nishta and Naik, A K},
  journal={Appl. Phys. Lett.},
  volume={113},
  number={11},
  pages={113101},
  year={2018},
  publisher={AIP Publishing},
  doi={https://doi.org/10.1063/1.5026775},
  url={https://pubs.aip.org/aip/apl/article-abstract/113/11/113101/35766/Tuning-of-geometric-nonlinearity-in-ultrathin?redirectedFrom=fulltext}
}

@book{lenzen1965development,
  title={Development of gravity pendulums in the 19th century},
  author={Lenzen, Victor Fritz and Multhauf, Robert P},
  volume={240},
  year={1965},
  publisher={Smithsonian Institution}
}

@article{kainz2018distortion,
  title={{Distortion-free measurement of electric field strength with a MEMS sensor}},
  author={Kainz, Andreas and Steiner, Harald and Schalko, Johannes and Jachimowicz, Artur and Kohl, Franz and Stifter, Michael and Beigelbeck, Roman and Keplinger, Franz and Hortschitz, Wilfried},
  journal={Nat. Electron.},
  volume={1},
  number={1},
  pages={68--73},
  year={2018},
  publisher={Nature Publishing Group UK London},
  doi={https://doi.org/10.1038/s41928-017-0009-5},
  url={https://www.nature.com/articles/s41928-017-0009-5}
}

@article{kareekunnan2021revisiting,
  title={Revisiting the mechanism of electric field sensing in graphene devices},
  author={Kareekunnan, Afsal and Agari, Tatsufumi and Hammam, Ahmed M M and Kudo, Takeshi and Maruyama, Takeshi and Mizuta, Hiroshi and Muruganathan, Manoharan},
  journal={ACS Omega},
  volume={6},
  number={49},
  pages={34086--34091},
  year={2021},
  publisher={ACS Publications},
  doi={https://doi.org/10.1021/acsomega.1c05530},
  url={https://pubs.acs.org/doi/10.1021/acsomega.1c05530}
}

@article{zhu2025non,
  title={{Non-Invasive Voltage Measurement Device Based on MEMS Electric Field Sensor and Applications}},
  author={Zhu, Xueqiong and Zhang, Ziyang and Hu, Chengbo and Wang, Zhen and Liu, Ziquan and Yang, Qing and Zhou, Jianglin and Qiu, Zhenhui and Bao, Shijie},
  journal={Electronics},
  volume={14},
  number={11},
  pages={2140},
  year={2025},
  publisher={MDPI},
  doi={https://doi.org/10.3390/electronics14112140},
  url={https://www.mdpi.com/2079-9292/14/11/2140}
}

@article{horenstein2001micro,
  title={{A micro-aperture electrostatic field mill based on MEMS technology}},
  author={Horenstein, Mark N and Stone, Patrick R},
  journal={J. Electrost.},
  volume={51-52},
  pages={515--521},
  year={2001},
  publisher={Elsevier},
  doi={https://doi.org/10.1016/S0304-3886(01)00048-1},
  url={https://www.sciencedirect.com/science/article/pii/S0304388601000481?via%3Dihub}
}

@article{peng2006design,
  title={Design and testing of a micromechanical resonant electrostatic field sensor},
  author={Peng, Chunrong and Chen, Xianxiang and Ye, Cao and Tao, Hu and Cui, Guoping and Bai, Qiang and Chen, Shaofeng and Xia, Shanhong},
  journal={J. Micromech. Microeng.},
  volume={16},
  number={5},
  pages={914},
  year={2006},
  publisher={IOP Publishing},
  doi={10.1088/0960-1317/16/5/006},
  url={https://iopscience.iop.org/article/10.1088/0960-1317/16/5/006}
}

@article{bahreyni2008analysis,
  title={Analysis and design of a micromachined electric-field sensor},
  author={Bahreyni, Behraad and Wijeweera, Gayan and Shafai, Cyrus and Rajapakse, Athula},
  journal={J. Microelectromechanical Syst.},
  volume={17},
  number={1},
  pages={31--36},
  year={2008},
  publisher={IEEE},
  doi={10.1109/JMEMS.2007.911870},
  url={https://ieeexplore.ieee.org/document/4447259}
}

@article{kobayashi2008microelectromechanical,
  title={{Microelectromechanical systems-based electrostatic field sensor using Pb(Zr,Ti)O$_3$ thin films}},
  author={Kobayashi, Takeshi and Oyama, Syoji and Takahashi, Masaharu and Maeda, Ryutaro and Itoh, Toshihiro},
  journal={Jpn. J. Appl. Phys.},
  volume={47},
  number={9S},
  pages={7533},
  year={2008},
  publisher={IOP Publishing},
  doi={10.1143/JJAP.47.7533},
  url={https://iopscience.iop.org/article/10.1143/JJAP.47.7533}
}

@article{yang2013design,
  title={{Design, fabrication and application of an SOI-based resonant electric field microsensor with coplanar comb-shaped electrodes}},
  author={Yang, Pengfei and Peng, Chunrong and Fang, Dongming and Wen, Xiaolong and Xia, Shanhong},
  journal={J. Micromech. Microeng.},
  volume={23},
  number={5},
  pages={055002},
  year={2013},
  publisher={IOP Publishing},
  doi={10.1088/0960-1317/23/5/055002},
  url={https://iopscience.iop.org/article/10.1088/0960-1317/23/5/055002}
}

@article{nan2013self,
  title={{Self-biased 215\,MHz magnetoelectric NEMS resonator for ultra-sensitive DC magnetic field detection}},
  author={Nan, Tianxiang and Hui, Yu and Rinaldi, Matteo and Sun, Nian X},
  journal={Sci. Rep.},
  volume={3},
  number={1},
  pages={1985},
  year={2013},
  publisher={Nature Publishing Group UK London},
  doi={https://doi.org/10.1038/srep01985},
  url={https://www.nature.com/articles/srep01985}
}

@article{li2017ultra,
  title={{Ultra-sensitive NEMS magnetoelectric sensor for picotesla DC magnetic field detection}},
  author={Li, Menghui and Matyushov, Alexei and Dong, Cunzheng and Chen, Huaihao and Lin, Hwaider and Nan, Tianxiang and Qian, Zhenyun and Rinaldi, Matteo and Lin, Yuanhua and Sun, Nian X},
  journal={Appl. Phys. Lett.},
  volume={110},
  number={14},
  pages={143510},
  year={2017},
  publisher={AIP Publishing},
  doi={https://doi.org/10.1063/1.4979694},
  url={https://pubs.aip.org/aip/apl/article/110/14/143510/594236/Ultra-sensitive-NEMS-magnetoelectric-sensor-for}
}

@article{xu2024subpicotesla,
  title = {Subpicotesla Optomechanical Magnetometry},
  author = {Xu, An-Ning and Li, Yifan and Li, Xiangliang and Liu, Bei and Liu, Yong-Chun},
  journal = {Phys. Rev. Lett.},
  volume = {133},
  issue = {15},
  pages = {153601},
  numpages = {6},
  year = {2024},
  month = {Oct},
  publisher = {American Physical Society},
  doi = {10.1103/PhysRevLett.133.153601},
  url = {https://link.aps.org/doi/10.1103/PhysRevLett.133.153601}
}

@article{middlemiss2016measurement,
  title={{Measurement of the Earth tides with a MEMS gravimeter}},
  author={Middlemiss, R P and Samarelli, Antonio and Paul, D J and Hough, James and Rowan, Sheila and Hammond, G D},
  journal={Nature},
  volume={531},
  number={7596},
  pages={614--617},
  year={2016},
  publisher={Nature Publishing Group UK London},
  doi={https://doi.org/10.1038/nature17397},
  url={https://www.nature.com/articles/nature17397}
}

@article{middlemiss2018microelectromechanical,
  title={Microelectromechanical system gravimeters as a new tool for gravity imaging},
  author={Middlemiss, Richard P and Bramsiepe, Steven G and Douglas, Rebecca and Hild, Stefan and Hough, James and Paul, Douglas J and Samarelli, Antonio and Rowan, Sheila and Hammond, Giles D},
  journal={Philos. Trans. R. Soc. A.},
  volume={376},
  number={2120},
  pages={20170291},
  year={2018},
  publisher={The Royal Society Publishing},
  doi={https://doi.org/10.1098/rsta.2017.0291},
  url={https://royalsocietypublishing.org/doi/10.1098/rsta.2017.0291}
}

@article{westphal2021measurement,
  title={Measurement of gravitational coupling between millimetre-sized masses},
  author={Westphal, Tobias and Hepach, Hans and Pfaff, Jeremias and Aspelmeyer, Markus},
  journal={Nature},
  volume={591},
  number={7849},
  pages={225--228},
  year={2021},
  publisher={Nature Publishing Group UK London},
  doi={https://doi.org/10.1038/s41586-021-03250-7},
  url={https://www.nature.com/articles/s41586-021-03250-7}
}

@article{fuchs2024measuring,
  title={Measuring gravity with milligram levitated masses},
  author={Fuchs, Tim M and Uitenbroek, Dennis G and Plugge, Jaimy and van Halteren, Noud and van Soest, Jean-Paul and Vinante, Andrea and Ulbricht, Hendrik and Oosterkamp, Tjerk H},
  journal={Sci. Adv.},
  volume={10},
  number={8},
  pages={eadk2949},
  year={2024},
  publisher={American Association for the Advancement of Science},
  doi={10.1126/sciadv.adk2949},
  url={https://www.science.org/doi/10.1126/sciadv.adk2949}
}

@article{yuan2015large,
  title={Large cooperativity and microkelvin cooling with a three-dimensional optomechanical cavity},
  author={Yuan, Mingyun and Singh, Vibhor and Blanter, Yaroslav M and Steele, Gary A},
  journal={Nat. Commun.},
  volume={6},
  number={1},
  pages={8491},
  year={2015},
  publisher={Nature Publishing Group UK London},
  doi={https://doi.org/10.1038/ncomms9491},
  url={https://www.nature.com/articles/ncomms9491}
}

@article{rej2025near,
  title={{Near-ground-state cooling in electromechanics using measurement-based feedback and a Josephson traveling-wave parametric amplifier}},
  author={Rej, Ewa and Cutting, Richa and Depellette, Joe and Datta, Debopam and Tiencken, Nils and Govenius, Joonas and Vesterinen, Visa and Liu, Yulong and Sillanp{\"a}{\"a}, Mika A},
  journal={Phys. Rev. Appl.},
  volume={23},
  number={3},
  pages={034009},
  year={2025},
  publisher={APS},
  doi={10.1103/PhysRevApplied.23.034009},
  url={https://journals.aps.org/prapplied/abstract/10.1103/PhysRevApplied.23.034009}
}

@article{liu2022quantum,
  title={Quantum backaction evading measurements of a silicon nitride membrane resonator},
  author={Liu, Yulong and Zhou, Jingwei and Mercier de L{\'e}pinay, Laure and Sillanp{\"a}{\"a}, Mika A},
  journal={New J. Phys.},
  volume={24},
  number={8},
  pages={083043},
  year={2022},
  publisher={IOP Publishing},
  doi={10.1088/1367-2630/ac88ef},
  url={https://iopscience.iop.org/article/10.1088/1367-2630/ac88ef}
}

@article{yuan2015silicon,
  title={Silicon nitride membrane resonators at millikelvin temperatures with quality factors exceeding $10^8$},
  author={Yuan, Mingyun and Cohen, Martijn A and Steele, Gary A},
  journal={Appl. Phys. Lett},
  volume={107},
  number={26},
  pages={263501},
  year={2015},
  publisher={AIP Publishing},
  doi={https://doi.org/10.1063/1.4938747},
  url={https://pubs.aip.org/aip/apl/article/107/26/263501/235945/Silicon-nitride-membrane-resonators-at-millikelvin}
}

@article{noguchi2016ground,
  title={{Ground state cooling of a quantum electromechanical system with a silicon nitride membrane in a 3D loop-gap cavity}},
  author={Noguchi, Atsushi and Yamazaki, Rekishu and Ataka, Manabu and Fujita, Hiroyuki and Tabuchi, Yutaka and Ishikawa, Toyofumi and Usami, Koji and Nakamura, Yasunobu},
  journal={New J. Phys.},
  volume={18},
  number={10},
  pages={103036},
  year={2016},
  publisher={IOP Publishing},
  doi={10.1088/1367-2630/18/10/103036},
  url={https://iopscience.iop.org/article/10.1088/1367-2630/18/10/103036}
}

@article{depellette2025amplitude,
  title={Amplitude noise cancellation of microwave tones},
  author={Depellette, Joe and Rej, Ewa and Herbst, Matthew and Cutting, Richa and Liu, Yulong and Sillanpää, Mika A.},
  journal={Rev. Sci. Instrum.},
  volume={96},
  number={8},
  pages={084705},
  year={2025},
  publisher={AIP Publishing},
  doi={https://doi.org/10.1063/5.0283567},
  url={https://pubs.aip.org/aip/rsi/article/96/8/084705/3360903/Amplitude-noise-cancellation-of-microwave-tones}
}

\clearpage
\pagebreak



\end{document}